\documentclass[
    ,final            
  ]
  {aipproc}

\layoutstyle{6x9}

\begin{document}

\title{The relationship between a topological Yang-Mills field 
and a magnetic monopole}


\author{Nobuyuki Fukui}
       {address={Department of Physics,
                 Graduate School of Science,
                 Chiba University, Chiba 263-8522, Japan}}

\author{Kei-Ichi Kondo}
       {address={Department of Physics,
                 Graduate School of Science,
                 Chiba University, Chiba 263-8522, Japan}}

\author{Akihiro Shibata}
       {address={Computing Research Center, KEK, Tsukuba  305-0801, Japan}}

\author{Toru Shinohara}
       {address={Department of Physics,
                 Graduate School of Science,
                 Chiba University, Chiba 263-8522, Japan}}

\begin{abstract}
We show that a Jackiw-Nohl-Rebbi solution, as the most general two-instanton,
generates a circular loop of magnetic monopole in four-dimensional Euclidean 
$SU(2)$ Yang-Mills theory.
\end{abstract}

\maketitle


\section{Introduction}
It is believed that a promising mechanism for quark confinement is
the dual superconductivity proposed in \cite{dualsuper,Polyakov77}.
In this mechanism, condensation of
magnetic monopoles causes confinement. Therefore, it must be shown
that magnetic monopoles to be condensed exist in Yang-Mills theory.
In the lattice simulation \cite{SKKISF09},
magnetic monopoles are exist in Yang-Mills theory
and magnetic monopole currents form loops.
In regard to this result, we can ask the following question.
\emph{Which configuration of the Yang-Mills field can be the source
for such magnetic monopoles?}
The simplest configuration examined first was the one-instanton configuration,
which is a solution of the self-dual equation $^\ast F=\pm F$
and has a unit instanton charge
$|Q_p|=1$. However, it has been confirmed in
\cite{KFSS08,CG95,BOT97}
that magnetic monopole loops are not generated from the one-instanton solution.

In this study,
we examine the Jackiw-Nohl-Rebbi (JNR) two-instanton solution.
We  demonstrate in a numerical way
that a circular loop is generated from the JNR solution.
We construct the magnetic monopole current
based on the nonlinear change of variables (NLCV)
and the reduction condition \cite{KSM08,KSSMKI08}.
The NLCV is a gauge-invariant extension of the Abelian projection
invented by 't Hooft \cite{tHooft81}
and enables one to extract magnetic monopoles
from the original Yang-Mills theory without breaking the gauge symmetry.
\section{The difinition of magnetic monopole}

We summarize the method
in a continuum $SU(2)$ Yang-Mills theory.
We introduce a color field ${\bf n}(x)$ with a unit length:
${\bf n}(x)=n^A(x)T^A, n^A(x)n^A(x)=1, (T^A :=\sigma_A/2)$,
where $\sigma_A\ (A=1,2,3)$ are Pauli matrices. 
The color field is determined by imposing the reduction condition.
It is given by minimizing the functional 
\begin{equation}
F_{\mbox{\footnotesize red}} 
:= \int d^4x \frac12 \mbox{tr} [\{D_\mu[{\bf A}]{\bf n}(x)\}^2   ] .
\end{equation}
The local minima are given by
the reduction differential equation (RDE) \cite{KFSS08}:
\begin{equation}
 - D_\mu[{\bf A}]D_\mu[{\bf A}]{\bf n}(x)=\lambda(x){\bf n}(x).\label{RDE}
\end{equation}

Once the RDE is solved for a given ${\bf A}_\mu$, 
we can obtain the gauge invariant magnetic monopole current $k^\mu$
by the following equations (NLCV).

\begin{equation}
 {\bf V}_\mu(x) :=
2 \mbox{tr}\left({\bf n}(x){\bf A}_\mu(x)\right)
{\bf n}(x)-ig^{-1}\left[\partial_\mu{\bf n}(x), {\bf n}(x)\right],
\label{V}
\end{equation}
\begin{equation}
 {\bf F}_{\mu\nu}[{\bf V}]
 =\partial_\mu{\bf V}_\nu
  -\partial_\nu{\bf V}_\mu
  -ig\left[{\bf V}_\mu, {\bf V}_\nu\right],
\end{equation}
\begin{equation}
 k^\mu(x)
 :=  \partial_\nu\!\,^\ast G^{\mu\nu}(x)\label{k}
 =  \epsilon^{\mu\nu\rho\sigma}\partial_\nu G_{\rho\sigma}(x)/2,\quad
 G_{\mu\nu}(x) = 2\mbox{tr}({\bf n}  {\bf F}_{\mu\nu}[{\bf V}]).
\end{equation}

We carry out this procedure numerically. We  use the lattice regularization and 
a lattice version of the NLCV \cite{SKS10} for numerical calculation.
In solving the RDE numerically, we must fix the asymptotic behavior of ${\bf n}$.
We recall that the instanton configuration approaches
a pure gauge at infinity:
$
 g{\bf A}_\mu(x)\rightarrow ih^\dagger(x)\partial_\mu h(x) + O(|x|^{-2}) .
\label{BehaviorOfA}$
Then,  ${\bf n}(x)$ as a solution of the reduction condition is supposed to
behave asymptotically 
$ {\bf n}(x)\rightarrow h^\dagger(x)T_3 h(x) + O(|x|^{-\alpha}) ,
\label{BehaviorOfn}$
for a certain value of $\alpha>0$. 
Actually, since we solve the RDE on a finite volume $V$,
we adopt a boundary condition as
${\bf n}^{{\mbox{\footnotesize bound}}}(x)=h^\dagger(x)T_3 h(x)
, \ x \in \partial V$.
\section{result}
\begin{figure*}[htbp]
  {\includegraphics[trim=0 0 -40 20, width=70mm]%
              {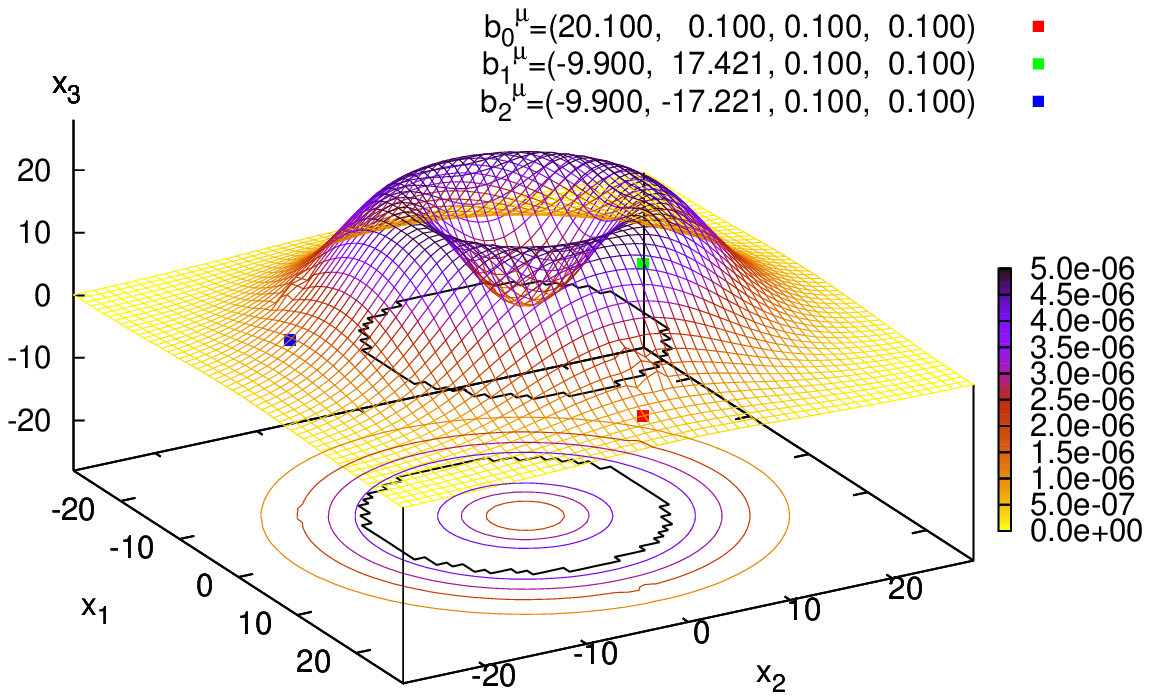}}%
  {\includegraphics[trim=50 30 0 0, width=65mm]%
                              {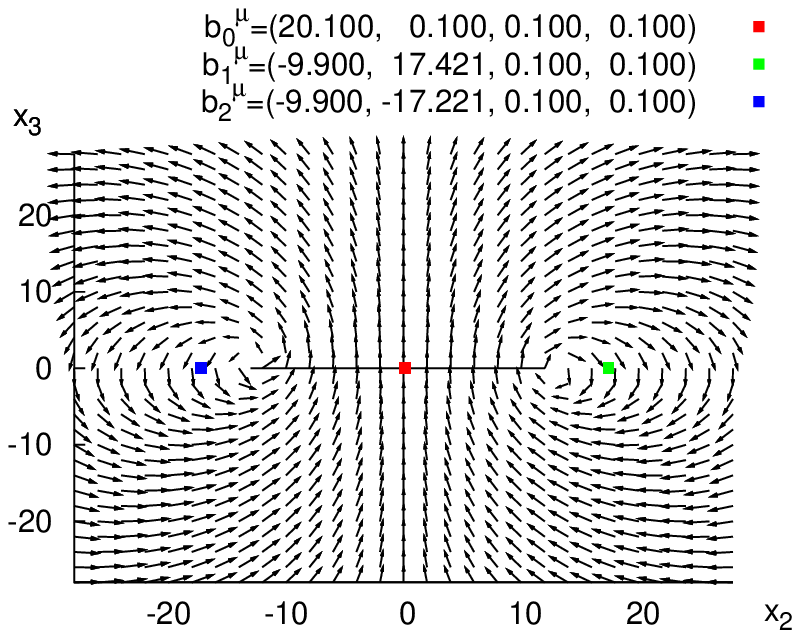}}
 \caption{(Left panel) The JNR two-instanton and the associated circular loop of
          the non-zero magnetic monopole current $k_\mu(x)$. The JNR 
          two-instanton is
          defined by fixing three scales $\rho_0=\rho_1=\rho_2=3$ and
          three pole positions $b_0^\mu,b_1^\mu,b_2^\mu$ which are arranged to
          be three vertices of an equilateral triangle.
          The grid shows an instanton charge density
          $D(x)$ on $x_1$-$x_2$ ($x_3=x_4=0$) plane.
          The black (thick) line on the base shows the magnetic monopole loop
          projected on $x_1$-$x_2$ plane,
          while colored (thin) lines on the base show the contour plot for
          the equi-$D(x)$ lines.
          (Right panel) The configuration of the color field
          ${\bf n}=(n_1,n_2,n_3)$ and a  circular loop
          of the magnetic monopole current $k_\mu(x)$
          obtained from the JNR two-instanton, 
          viewed in the $x_2$-$x_3$ ($x_1=x_4=0$) plane which is off three poles.
          The magnetic monopole current and the three poles
          of the JNR solution are projected on the same plane.
          Here the $SU(2)$ color field $(n_1,n_2,n_3)$ is identified
          with a unit vector in the three-dimensional space $(x_1,x_2,x_3)$.
}
 \label{fig:monopole}
\end{figure*}
The explicit form of the JNR two-instanton solution is 
\begin{equation}
 g{\bf A}_\mu(x) =T^A \eta_{\mu\nu}^{A(-)}
                  \phi_{\rm JNR}^{-1}
                  \sum_{r=0}^2\frac{2\rho_r^2\left(x^\nu-b^\nu_r\right)}
                                   {(|x-b_r|^2)^2},\ 
\phi_{\rm JNR} := \sum_{r=0}^2\frac{\rho_r^2}{|x-b_r|^2},
\end{equation}
where $|x|^2=x_\mu x^\mu$.
The JNR two-instanton is specified by three pole positions
$(b_0^1,b_0^2,b_0^3,b_0^4)$,
$(b_1^1,b_1^2,b_1^3,b_1^4)$,
$(b_2^1,b_2^2,b_2^3,b_2^4)$ 
and three scale parameters $\rho_0,\rho_1,\rho_2$.
The result for the particular set of parameters
is shown in FIGURE \ref{fig:monopole}.
The following two points are notable \cite{NKSS10}.
\begin{itemize}
 \item Non-zero monopole currents originating from JNR two-instanton form
       a circular loop located  near the maxima of the instanton charge density
       (Left panel).
 \item ${\bf n}$ field is winding around the loop
       and indeterminate at points where the loop pass (Right panel).
       The configurations of the color field giving the magnetic monopole
       loop were made available for the first time in this study
       based on the NLCV. 
\end{itemize}

\section{conclusion and discussion}
For the JNR two-instanton solution,
we have solved the RDE in a numerical way and obtained the magnetic monopole
currents and
discovered that non-zero magnetic monopole currents form a circular loop which
is located near the maxima of the instanton charge density. 
In our previous work
\cite{KFSS08},
we have found the two-meron solution, which is a solution of the classical
Yang-Mills equation with a unit \emph{total}
topological charge $|Q_p|=1$, leads to a
circular loop of magnetic monopole in an analytical way.
Combining these results,
we have found that both the JNR solution and two-meron solution with same
asymptotic behavior at infinity
${\bf A}_\mu(x)\sim O(|x|^{-1}),\ |x|\rightarrow\infty$
generate circular loops of magnetic monopole.
We expect that this loop is responsible for confinement
in the dual superconductor picture.


\begin{theacknowledgments}
This work is supported by AGSST of Chiba University and the Grant-in-Aid
for Scientific Research (C) 21540256 from JSPS.
\end{theacknowledgments}



\bibliographystyle{aipproc}   

\bibliography{sample}

\IfFileExists{\jobname.bbl}{}
 {\typeout{}
  \typeout{******************************************}
  \typeout{** Please run "bibtex \jobname" to optain}
  \typeout{** the bibliography and then re-run LaTeX}
  \typeout{** twice to fix the references!}
  \typeout{******************************************}
  \typeout{}
 }

\end{document}